\documentclass[a4paper,12pt]{article}

\usepackage{amssymb}
\usepackage[numbers,sort&compress]{natbib}

\def\bea{\begin{eqnarray}}
\def\eea{\end{eqnarray}}
\def\be{\begin{equation}}
\def\ee{\end{equation}}
\def\ba{\begin{array}}
\def\ea{\end{array}}
\def\nn{\nonumber}

\def \lsim{\mathrel{\vcenter
{\hbox{$<$}\nointerlineskip\hbox{$\sim$}}}}
\def \gsim{\mathrel{\vcenter
{\hbox{$>$}\nointerlineskip\hbox{$\sim$}}}}

\begin{document}
\baselineskip=16.6pt
\pagestyle{plain}

\renewcommand{\theequation}{\arabic{section}.\arabic{equation}}
\setcounter{page}{1}

\setlength\arraycolsep{2pt}

\begin{titlepage} 

\rightline{\footnotesize{DESY 08-059}} \vspace{-0.2cm}
\rightline{\footnotesize{CERN-PH-TH/2008-107}} \vspace{-0.2cm}
\rightline{\footnotesize{ZMP-HH/08-9}} \vspace{-0.2cm}

\begin{center}

\vskip 0.4 cm

{\LARGE  \bf Constraints on modular inflation in\\[2mm] supergravity and string theory}

\vskip 0.7cm

{\large 
Laura Covi$^{a}$, Marta Gomez-Reino$^{b}$, Christian Gross$^{c}$, \\[1mm]
Jan Louis$^{c,d}$, Gonzalo A. Palma$^{a}$, Claudio A. Scrucca$^{e}$
}

\vskip 0.5cm

{\it
$^{a}$Theory Group, Deutsches Elektronen-Synchrotron DESY,\\ 
Notkestrasse 85, D-22603 Hamburg, Germany\\
$^{b}$Theory Division, Physics Department, CERN, \\
CH-1211 Geneva 23, Switzerland\\
$^{c}$II. Institut f\"ur Theoretische Physik, Universit\"at Hamburg, \\
Luruper Chaussee 149, D-22761 Hamburg, Germany\\
$^{d}$Zentrum f\"ur Mathematische Physik,
Universit\"at Hamburg,\\
Bundesstrasse 55, D-20146 Hamburg, Germany\\
$^{e}$Inst. de Th. des Ph\'en. Phys.,
Ecole Polytechnique F\'ed\'erale de Lausanne, 
\mbox{CH-1015 Lausanne, Switzerland}\\
}

\vskip 0.8cm

\end{center}

We perform a general algebraic analysis on the possibility of realising 
slow-roll inflation in the moduli sector of string models. 
This problem turns out to be very closely related to the characterisation 
of models admitting metastable vacua with non-negative cosmological constant. 
In fact, we show that the condition 
for the existence of viable inflationary trajectories is a deformation 
of the condition for the existence of metastable de Sitter vacua. 
This condition depends on the ratio between the scale of inflation and the 
gravitino mass and becomes stronger as this parameter grows. 
After performing a general study within arbitrary supergravity models, 
we analyse the implications of our results in several examples. 
More concretely, in the case of heterotic and orientifold string 
compactifications on a Calabi-Yau in the large volume limit we show 
that there may exist fully viable models, allowing both for inflation 
and stabilisation. Additionally, we show that subleading corrections 
breaking the no-scale property shared by these models always allow 
for slow-roll inflation but with an inflationary scale suppressed with 
respect to the gravitino scale. A scale of inflation larger than the 
gravitino scale can also be achieved under more restrictive circumstances 
and only for certain types of compactifications.
\end{titlepage}

\newpage

\section{Introduction} \setcounter{equation}{0}

Our current understanding of the very early universe is consistent with 
a period of dramatic accelerated expansion known as cosmological 
inflation~\cite{Guth:1980zm}. 
The simplest and, so far, most successful way of modelling this stage
consists in the `slow-roll' motion of a single scalar field --the inflaton-- 
behaving as a perfect fluid with negative pressure driving the 
universe into accelerated expansion~\cite{Linde:1981mu, Albrecht:1982wi}
(for a recent review, see \cite{Lyth:1998xn}).
A crucial ingredient in these types of models (generically referred to 
as slow-roll inflation) is the flatness of the scalar field potential 
characterising the inflaton's dynamics. 
On the one hand, this feature allows for inflation to last long enough, 
so that the universe can become flat, homogeneous and isotropic at 
cosmological scales \cite{Guth:1980zm, Linde:1981mu}. 
On the other, it is required to obtain the correct prediction for the 
spectrum of primordial density fluctuations~\cite{Mukhanov:1981xt,Lyth:1998xn} 
as observed in precision 
measurements of the cosmic microwave background~\cite{Spergel:2006hy} 
and large scale structure of the 
universe~\cite{Tegmark:2006az, Sanchez:2005pi}. 

Despite of its simplicity, a completely satisfactory realisation of 
slow-roll inflation in supergravity and string theory has remained 
elusive, the main reason for this being the difficulty of 
ensuring the flatness of the inflaton potential \cite{Copeland:1994vg}. 
Up to date, the most popular strategy to achieve inflation in string 
theory has consisted in the search of suitable inflationary trajectories
within the vast landscape of string vacua, by studying the class of 
$\mathcal N = 1$ supergravity models arising from string theory. 
The large amount of freedom available in string compactifications, 
such as that coming from fluxes, torsion and/or non-perturbative effects, 
suggests that there should be no obstacles in obtaining a rich variety 
of scalar potentials, even possessing flat directions. 
However, in early attempts to achieve inflation, it was already 
understood that there are actually severe restrictions towards this possibility,
particularly for the identification of the inflaton within the moduli 
sector~\cite{Binetruy:1986ss, Brustein:1992nk, Banks:1995dp}.
In practise most of the successful scenarios of string inflation involve 
an additional sector beyond the moduli like, for instance, the uplifting sector used in
most of the constructions of de Sitter (dS) vacua 
with fixed moduli. Examples of this type are models of inflation 
based on the KKLT scenario~\cite{Kachru:2003aw} where the joint 
contribution of non-perturbative effects and an explicit supersymmetry 
breaking term induced by anti-D3 branes allows to have dS 
vacua with a stable volume-modulus. Recently many interesting 
examples of such models of modular inflation have been 
proposed~\cite{BlancoPillado:2004ns, Simon:2006du, Lalak:2005hr, Conlon:2005jm, BlancoPillado:2006he, Bond:2006nc, de Carlos:2007dp, Lalak:2007vi, Grimm:2007hs, Linde:2007jn}. 
Moreover, with the advent of new uplifting mechanisms replacing the 
one offered by anti-D3 branes, new string-inspired realisations of 
inflation with similar characteristics have been constructed~
\cite{Westphal:2005yz, Ellis:2006ara, Brax:2007fe}. Also some progress 
has been made recently towards a more general understanding of the origin 
of the difficulties in realising inflation in the moduli sector of 
string compactifications~\cite{Badziak:2008yg, BenDayan}.

As a matter of fact, the problem of finding viable models of single field
slow-roll inflation turns out to be closely related to the problem of 
finding metastable vacua. A considerable step forward in the last question 
has been the new understanding of the circumstances under which 
metastable Minkowski vacua may exist in supergravity models. 
In ref.~\cite{GomezReino:2006dk} for instance, it was shown that positivity of 
the scalar mass matrix along the direction associated to the scalar partners 
of the would be Goldstino (sGoldstinos) implies a strong necessary condition 
on the K\"ahler potential $K$ independently of the superpotential $W$.
This condition was shown to have strong implications when applied to the 
moduli sector of the simplest string compactifications available 
(see also \cite{Brustein:2004xn, Lebedev:2006qq, GomezReino:2006wv, Lust:2006zg, GomezReino:2007qi}). 
More recently, in ref.~\cite{Covi:2008ea}, a more comprehensive study of 
the requirements for the existence of dS vacua in the moduli sector of 
Calabi-Yau string compactifications was carried out. 
From these studies it emerged again that the crucial quantity controlling 
metastability is the value of the mass matrix along a special direction 
defined by the Goldstino vector, which depends only on $K$.
It was shown, in fact, that the value of the mass matrix along any other 
direction can be made arbitrarily large by appropriately choosing $W$, 
and that, once a suitable choice of $K$ is adopted, it is always possible 
to construct metastable de Sitter vacua as long as there is enough freedom 
to tune the superpotential of the model. 

The purpose of this work is to show that a similar analysis can be performed 
to determine if a supergravity model may possess flat-enough directions 
allowing for inflation.
We find that the resulting condition is a generalisation of the constraint 
granting the existence of metastable dS vacua, and that, as much as for the 
realisation of these types of vacua, the main obstruction towards the 
realisation of inflation comes from the choice of the K\"ahler potential $K$. 
There exists, nevertheless, a significant difference between these two 
situations. At the point where the final stabilisation of the moduli occurs, 
the value of $V$ is related to the cosmological constant 
$\Lambda$ by $V = \Lambda$, which is tiny. In particular, one certainly has 
phenomenologically $V \ll m_{3/2}^2 M_{\mathrm{Pl}}^{2}$.
During inflation, on the other hand, the value of $V$ is related to
the Hubble constant by $H^2 \simeq V/ (3 M_{\mathrm{Pl}}^{2})$, and the
gravitino mass generically differs from the one at the stabilised
vacuum. We expect, however, the order of magnitude of that mass to 
remain the same in most cases, unless some extra tuning is enforced 
in the model.
We therefore consider in the following a single scale for the gravitino
mass, and such situation is surely realised for example in models where
inflation is driven by the large F-term of a field whose contribution to
$W$ is nevertheless suppressed.

In general, it is then desirable to have $H \gg m_{3/2}$, since the scale
of inflation should be much higher than the electroweak scale and particle
phenomenology calls for $m_{3/2}$ comparable to, or lower than, that scale.
As already noticed in \cite{Covi:2008ea}, the condition for achieving 
a massless sGoldstino becomes stronger as the ratio 
$V/ (m_{3/2}^2 M_{\mathrm{Pl}}^{2})$ increases. 
This means in particular 
that, in generic supergravity models, the condition for getting slow roll 
inflation is stronger than the condition for realising moduli stabilisation. 
The difference between these two situations can be conveniently parametrised 
in terms of the following quantity
\be
\gamma =\frac13 \, \frac {V}{m_{3/2}^2 M_{\mathrm{Pl}}^{2}}  \simeq  \frac{H^{2}}{m_{3/2}^2}\,.
\label{gam}
\ee
Let us emphasise however that the results presented in this work are also valid for models 
in which the gravitino mass changes strongly between inflation and the present vacuum. In such 
cases, the parameter $\gamma $ has to be rescaled accordingly, 
and the comparison between the values of $\gamma $ during and after inflation 
is still a useful indication of the difficulty in realising the scenarios.

The outline of this paper is as follows. In Section 2 we derive the condition
that a generic supergravity theory must fulfil in order to allow for 
slow-roll inflation. In Section 3 we illustrate our results through some 
simple examples. In Section 4 we apply it to the case of heterotic and 
orientifold string compactifications on a Calabi-Yau in the large volume 
limit, and see what kind of information can be extracted. 
In Section 5 we study the effect of subleading corrections to the K\"ahler 
potential breaking the no-scale property shared by the models presented 
in Section 4 and its implications on the inflationary analysis. 
Finally in Section 6 we present our conclusions.

\section{Slow-roll inflation in supergravity} \setcounter{equation}{0}

To begin with, let us consider a generic model involving several complex 
neutral scalar fields $\phi^i$, with a Lagrangian of the type 
(in Planck units $M_{\mathrm{Pl}} = 1$):
\be
{\cal L} = \frac{1}{2}  R - g_{i \bar \jmath} \, \partial \phi^{i} \partial \bar \phi^{\bar \jmath} 
- V(\phi^i,\bar \phi^{\bar \imath}) \,.
\label{s1: act}
\ee
The metric $g_{i \bar \jmath}$ must be Hermitian and positive definite, but is otherwise arbitrary.
The realisation of a successful and viable stage of slow-roll inflation in such a model requires the 
existence of a region in field space where the potential in the canonical basis is sufficiently flat.
In the case of a single real field, this corresponds to the requirement of having small slow-roll parameters
$\epsilon$, $|\eta|\ll1$, where 
\bea
\epsilon = \frac{1}{2} \left( \frac{V'}{V} \right)^{2}, \qquad  \eta = \frac{V''}{V} \, ,   \label{slow-roll-1}
\eea
with $'$ denoting derivatives with respect to the canonically normalised field.
These conditions get modified in the multi-field case.
At lowest order in the slow-roll approximation the trajectory in field-space along which 
inflation is realised is given by the direction $|\nabla V|^{-1}\nabla_{i} V$,
where $|\nabla V| = \sqrt{\nabla^j V \nabla_j V}$,
whereas deviations from this direction are controlled by the tensor:
\begin{eqnarray}
N^I_{\;\;J} = \frac{1}{V} \left(\matrix{
\displaystyle{\nabla^i \nabla_j V} \!&\! \displaystyle{ \nabla^i \nabla_{\bar \jmath} V} \nn \\
\displaystyle{\nabla^{\bar \imath} \nabla_j V} \!&\! \displaystyle{\nabla^{\bar \imath} \nabla_{\bar \jmath} V}}
\hspace{-35pt}\right) \,  ,
\end{eqnarray}
where $I=(i,\bar \imath)$ and $J = (j,\bar \jmath)$, and $\nabla_{i}$ denotes a derivative which is covariant with
respect to the metric $g_{i \bar \jmath}$. Then, the generalised version of the slow-roll parameters (\ref{slow-roll-1})
can be defined in the following way~\cite{Burgess:2004kv}:
\begin{eqnarray}
\epsilon &=& \frac {\nabla^i V \nabla_i V}{V^2} \,,  \\
\eta &=& \mbox{min eigenvalue} \left\{ N \right\}\,.    \label{slowr}
\end{eqnarray}
Let us mention here that a strict characterisation of the slow-roll conditions would
require us to distinguish between dynamical effects parallel and perpendicular to the inflaton's
trajectory~\cite{GrootNibbelink:2000vx}. 
In particular, $\eta$ given in eq.~(\ref{slow-roll-1}) would have to be generalised in
such a way that it coincides with the projection 
$\eta_{||}$ of $N$ along the direction $|\nabla V|^{-1}\nabla_{i} V$.\footnote{Also 
a second slow roll-parameter $\eta_{\bot}$, depending on $N$, may be defined~\cite{GrootNibbelink:2000vx}. 
Loosely speaking, $\eta_{\bot}$ depends only on those elements of $N$ mixing the tangent vector
$|\nabla V|^{-1}\nabla_{i} V$ with the normal vector relative to the inflaton trajectory.} 
Notice from the definition (\ref {slowr}), 
however, that for any given unit vector $u^I=(u^i,u^{\bar \imath})$ the following inequality is always satisfied:
\bea
\eta \le u_I N^I_{\;\; J} u^J. \label{eta-uMu}
\eea
Indeed, one can always decompose $u^I$ as 
$u^I = \sum_k c_{(k)} \omega^I_{(k)}$, where the $\omega^I_{(k)}$'s represent a 
basis of orthogonal and normalised eigenvectors of $N$ with eigenvalues $\lambda_{(k)}$. 
Since the $u^I$'s 
are unit vectors, the coefficients $c_{(k)}$ satisfy $\sum_k |c_{(k)}|^2 = 1$ and so it immediately  follows that
$u_I N^I_{\;\;J} u^J = \sum_k |c_{(k)}|^2 \lambda_{(k)} \ge {\rm min}\{\lambda_{(k)}\} = \eta$.
In particular, one finds that $\eta \le \eta_{||}$. Nevertheless, in 
order to avoid significant levels of isocurvature perturbations, 
a phenomenologically successful model of inflation requires the projection 
of $N$ along directions perpendicular to $|\nabla V|^{-1}\nabla_{i} V$ to be much larger 
than $\eta_{||}$. This means that $\eta \simeq \eta_{||}$, as contributions to $\eta$ coming 
from projecting $N$ along directions perpendicular to $|\nabla V|^{-1}\nabla_{i} V$ 
have to be suppressed. 

Let us consider now the situation in a generic supergravity theory involving only chiral multiplets.
Recall that in supergravity the two-derivative Lagrangian can be written in terms of the real function 
$G = K + \log |W|^2$ and its derivatives with respect to the chiral multiplets $\Phi^i$ (and their 
conjugates $\bar \Phi^{\bar \jmath}$) which are denoted by lower indices $i$ (and $\bar \jmath$). 
The kinetic term of the scalar fields involves the K\"ahler metric $g_{i \bar \jmath} = G_{i \bar \jmath}$, 
which can be used to raise and lower indices and depends only on $K$. The K\"ahler metric is 
assumed to be positive definite and defines a K\"ahler geometry for the manifold spanned by the 
scalar fields. The scalar potential for this kind of theories takes the following simple form:
\begin{eqnarray}
V = e^{G} \big(G^{i} G_{i} - 3  \big)\,. 
\label{s2:F-term-G} 
\end{eqnarray}
The auxiliary fields of the chiral multiplets are fixed by their equations of motion to be $F^i=m_{3/2}G^i$
with a scale set by the gravitino mass $m_{3/2}=e^{G/2}$. Whenever $F^i\neq 0$ at the vacuum
supersymmetry is spontaneously broken, and the direction $G^i$ in the space of chiral fermions 
defines the Goldstino which is absorbed by the gravitino in the process of supersymmetry breaking.
The unit vector defining this direction is given by:
\be
f_i = \frac {G_i}{\sqrt{G^j G_j}} \,.
\label{gold}
\ee
Note that such direction can be different during and after inflation.

In these theories achieving small values for $\epsilon$ and $\eta$ is not 
completely trivial. 
This is due to the fact that the potential $V$ is constrained to be a 
specific function of $K$ and $W$, and is therefore not entirely arbitrary. 
Nevertheless, if $K$ is appropriately chosen, it is always possible to make 
$\epsilon$ and $\eta$ arbitrarily small by tuning $W$. 
To see this, we must first compute the first and second derivatives 
of $V$ and express them in terms of the parameters of the theory. 
These are:
\bea
\hspace{-.5cm}\nabla_i V &=& e^G \Big(G_i + G^j \nabla_i G_j \Big) + G_i V \,,\label{vi} \\
\hspace{-.5cm}\nabla_i \nabla_{\bar \jmath} V &=&  e^G \Big(g_{i \bar \jmath} + 
\! \nabla_i G_k \nabla_{\bar \jmath} G^k \!
- R_{i \bar \jmath p \bar q} G^p G^{\bar q}\Big) \! + G_i \nabla_{\bar \jmath} V \! + G_{\bar \jmath} \nabla_i V 
\!+\! \big(g_{i \bar \jmath} - G_i G_{\bar \jmath} \big) V \, ,  \quad \label{vij} \\
\hspace{-.5cm}\nabla_i \nabla_{j} V &=&  e^G \Big( 2 \nabla_{i} G_{j} +  G^k \nabla_i \nabla_j G_k  \Big) \! + G_i \nabla_{j} V \! + G_{j} \nabla_i V 
\!+\! \big(\nabla_{i} G_{j} - G_i G_{j} \big) V \, .  \label{vij-non-diag}
\eea
Notice that $G_{i}$, $\nabla_{i} G_j$ and $\nabla_{i} \nabla_{j} G_k$ depend 
on the superpotential and more precisely on 
$(\log W)_i$, $(\log W)_{ij}$ and $(\log W)_{ijk}$, which are independent 
quantities. This means that $W$ may be varied in an arbitrary way 
in order to adjust $\nabla_i V$ and $N$.
It is clear then that, for a given K\"ahler potential, it is always 
possible to make $\epsilon$ arbitrarily small, simply by tuning 
$G^{k} \nabla_{i} G_{k}$ with respect to $G_{i}$ in eq.~(\ref{vi}). 
On the other hand, to achieve a small $|\eta|$, we need to have sufficient 
control on the entries of the matrix $N$. Observe that by tuning 
$\nabla_i \nabla_j G_k$ it is possible to set $\nabla_i \nabla_j V$ to 
any desired value, and the quantities $\nabla_i G_j$ to make most of the 
eigenvalues of $\nabla_i \nabla_{\bar \jmath} V$ large and positive. 
The only restriction comes from the fact that the projection of 
$\nabla_i \nabla_{\bar \jmath} V$ along the Goldstino direction 
(\ref{gold}) is actually constrained by eq.~(\ref{vi}) 
(which has already been fixed to make $\epsilon$ small) and therefore cannot 
be adjusted so easily.  Nevertheless, if the choice 
of $K$ allows for it, one can still make this last direction flat enough 
by tuning the remaining quantities $G_{i}$.

From the previous discussion it remains to be learned under which 
circumstances a given $K$ is suitable to produce such a flat direction. 
To find this out, let us recall that eq.~(\ref{eta-uMu}) is valid for any unit 
vector $u$.
We can then derive an upper bound on $\eta$ for the particular 
choice $u_I=(e^{-i \alpha} f_i,e^{i \alpha} f_{\bar \imath})/\sqrt{2}$, 
$u^J=(e^{i \alpha} f^j,e^{-i \alpha} f^{\bar \jmath})/\sqrt{2}$, which is
associated to the Goldstino direction $f^{i}$ given in eq.~(\ref{gold}):\footnote{
Notice that if  $G^{k} \nabla_{i} G_{k} \propto G_i $ then $V_i \propto G_i$ and the inflaton and Goldstino directions 
in field space are aligned. Then the
value of $\eta_{||}$ is equal to the right hand side of (\ref{eta-Vff-alpha}) with $\alpha = 0$.} 
\bea
\eta \le \frac {\nabla_i \nabla_{\bar \jmath} V}V f^i f^{\bar \jmath} 
+ {\rm Re}\, \Bigg\{e^{2 i \alpha} \frac {\nabla_i \nabla_j V}V f^i f^j \Bigg\} \,. \label{eta-Vff-alpha}
\eea
Averaging this over the two orthogonal choices $\alpha=0,\pi/2$ one finally deduces the following 
simple bound, depending only on the Hermitian block of the Hessian matrix:
\bea
\eta \le \frac {\nabla_i \nabla_{\bar \jmath} V}V f^i f^{\bar \jmath} \,. \label{eta-constr}
\eea
Using now eq.~(\ref{vij}) it is straightforward to find:
\bea
\frac {\nabla_i \nabla_{\bar \jmath} V}V f^i f^{\bar \jmath} = - \frac 23 
+ \frac 4{\sqrt{3}} \frac 1{\sqrt{1+\gamma}} {\rm Re} \Bigg\{\frac {\nabla_i V}{V} f^i \Bigg\}
+ \frac \gamma{1 + \gamma} \frac {\nabla^i V \nabla_i V}{V^2} + \frac {1 + \gamma}\gamma 
\hat\sigma (f^i)\,,  \qquad  \label{vijff}
\eea
where the parameter $\gamma$ is given by eq.~(\ref{gam}) and the function $\hat\sigma(f^i)$ is 
defined to be
\be
\hat\sigma (f^i) = \frac 23 - R(f^i)\,,
\label{shat}
\ee
where $R(f^i) = R_{i \bar \jmath p \bar q} \, f^i f^{\bar \jmath} f^p f^{\bar q}$ 
denotes the holomorphic sectional curvature along the Goldstino direction $f^i$. 
Note that the quantity $\hat \sigma$ is the normalised version of the 
homogeneous quantity $\sigma$ that was introduced in ref.~\cite{Covi:2008ea}:\footnote{ In the notation of
ref.~\cite{Covi:2008ea} the bound (\ref{eta-constr}) takes the form 
$\eta  \le e^{G} \lambda/(V G^{i} G_{i})$, where 
$\lambda  =  - 2/3 \, e^{-G} V \big(e^{-G} V  + 3 \big) 
+ \sigma  +  2 e^{-G} (G^{m}  V_{m} +G^{\bar n} V_{\bar n}) + V^{n}V_{n} $.}
$\hat\sigma (f^i)= \sigma(G^i)/(G^k G_k)^2$.

Since $f^i$ is a unit vector, it is clear that $|f^i \nabla_i V/V| \le \sqrt{\epsilon}$. 
Using this inequality, the definition of $\epsilon$, and the result given in (\ref{vijff})
we finally obtain the following simple upper bound on $\eta$:
\be\label{ub}
\eta \le \eta_{\rm max} \equiv - \frac 23 + \frac 4{\sqrt{3}} \frac 1{\sqrt{1+\gamma}} \sqrt{\epsilon}
+ \frac \gamma{1+\gamma} \epsilon + \frac {1 + \gamma}\gamma \hat\sigma(f^i) \,.
\ee
Notice now that $\eta_{\rm max}$ should be either negative and very small or positive ($\eta_{\rm max} \gsim 0$) 
in order for the bound (\ref{ub}) to be compatible with the requirement of having a small $|\eta|$. 
More precisely, assuming $\epsilon \ll 1$, one needs:
\be
\hat\sigma (f^i) \gsim \frac 23 \frac \gamma{1+\gamma} \,.
\label{cond}
\ee
This condition can also be rewritten in the following alternative form, which has the same structure 
as the conditions derived in refs.~\cite{GomezReino:2006dk,GomezReino:2006wv,GomezReino:2007qi}:
\be
R(f^i) \lsim \frac 23 \frac 1{1 + \gamma} \,.
\label{condalt}
\ee
The condition (\ref{cond}), or equivalently (\ref{condalt}), represents our main result and implies a strong 
restriction on the K\"ahler potential, generalising the one obtained in refs.~\cite{Badziak:2008yg,BenDayan}
for single fields models. If it is satisfied, one still needs to further 
tune the superpotential to adjust $\eta$ to a sufficiently small value 
compatible with current data.

For $\gamma \ll 1$, this condition reduces to $\hat\sigma(f^i) \gsim 0$ (or $R(f^i) \lsim 2/3$), 
which coincides with the condition for the existence of metastable dS vacua with small cosmological constant. 
On the other hand, for $\gamma \gg 1$, it tends to the more restrictive condition $\hat\sigma(f^i) \gsim 2/3$ (or $R(f^i) \lsim 0$). 
Since $\gamma = (H/m_{3/2})^2$ parametrises the ratio between the 
Hubble scale $H$ and the gravitino scale $m_{3/2}$, this means that 
inflationary scales much smaller than the gravitino scale are as difficult 
to realise as dS vacua, whereas higher inflationary scales are instead more 
difficult to realise.

One can study the implications of the condition (\ref{condalt}) exactly in the same way as was done in 
refs.~\cite{GomezReino:2006dk,GomezReino:2006wv,GomezReino:2007qi}. In particular, one can 
derive a constraint involving only the K\"ahler potential by minimising 
the sectional curvature with respect to the variables $f^i$, taking into account that these variables 
are normalised to one: $f^if_i=1$. This implies the condition that the minimal value of the sectional 
curvature $R_{\rm min}$ should be less than $2/[3(1+\gamma)]$. Moreover, if $R_{\rm min}$ 
satisfies that bound, the direction $f^i$ is then 
constrained to lie within a cone centred around the particular direction 
that minimises the sectional curvature. 
This procedure can be performed explicitly for particular classes of models, 
like for instance those for which the scalar manifold factorises into a product of one-dimensional scalar 
manifolds or also for coset scalar manifolds. More precisely, for factorisable manifolds it is easy to 
show that the sectional curvature satisfies a lower bound in terms of the scalar curvatures $R_i$ of the 
one-dimensional submanifolds which is given by: $R(f^i) \ge (\sum_i R_i^{-1})^{-1}$. 
For coset manifolds, on the other hand, the Riemann tensor has a very special structure. 
One can show that in those cases the sectional curvature turns out to be constant and to depend 
only on some overall curvature scale $R_{\rm all}$, which depends on the particular coset 
manifold being considered: $R(f^i) = R_{\rm all}$ (see \cite{GomezReino:2006wv} for more details). 

It is worth pointing out that the presence of vector multiplets gauging isometries of the chiral multiplet
geometry can quantitatively change the right-hand side of the constraint (\ref{condalt}). 
This is mainly due to the fact that the $D$-term contributions to the scalar potential are positive 
definite. More precisely, for a given value of the potential $V$, increasing the ratio between 
the $D$-term and the $F$-term contributions to the potential has the net effect of 
reducing the left-hand side of (\ref{condalt}) and therefore making the constraint milder 
\cite{GomezReino:2007qi}. This could be used to partly compensate the strengthening of the condition 
induced by increasing $\gamma$.
A more radical improvement of the situation can be obtained by relying on genuine 
constant Fayet-Iliopoulos terms \cite{Binetruy:1996xj}. However, this possibility is severely 
constrained within supergravity, and implies a rather peculiar gauging of the $R$-symmetry,
which does not seem to emerge in any kind of string construction~\cite{Binetruy:2004hh}.

\section{Simple examples}\setcounter{equation}{0}

The simplest example one can study is the case of supergravity models involving a single 
chiral superfield with a canonical K\"ahler potential:
\be
K = \bar X X \,.
\ee
For this scalar manifold the Riemann tensor vanishes. From (\ref{shat})
we get then that $\hat\sigma = 2/3$, and the condition (\ref{cond}), or equivalently 
(\ref{condalt}), can always be satisfied independently of the value of $\gamma$. This implies in 
particular that there is no obstruction in this case to build a model with any scale of inflation. 
In models with several fields of this type, that is, with $K=\sum_i\bar X^iX^i$, the components 
of the Riemann tensor will also vanish and therefore the situation is exactly the same.

Another simple case that can be studied is the case of a field with a logarithmic K\"ahler potential, for which a no-go theorem is discussed 
in \cite{Badziak:2008yg, BenDayan}:
\be\label{ene}
K = - n \log \big(T + \bar T \big) \,,
\ee
which governs the dynamics of moduli fields arising in simple examples of string compactifications. 
The one-dimensional scalar manifold has in this case a constant sectional curvature which is 
simply given by $R = 2/n$. From here we get that $\hat\sigma = 2/3(1-3/n)$. This means that 
the condition (\ref{cond}), or (\ref{condalt}), can be satisfied only if 
$$
n \gsim 3(1 + \gamma)\,.
$$
It is then clear that a model with $\gamma \gg 1$ cannot be 
built within this setup, as $n$ is typically a number of order $1$. For instance the overall 
K\"ahler modulus in string models has $n=3$ and thus, even including subleading corrections to 
the K\"ahler potential, one can at best achieve a small $\gamma$ of the order of the subleading 
corrections.\footnote{In ref.~\cite{Badziak:2008yg} a model of this kind 
is proposed where a sizable $\gamma$ is achieved by going to a regime 
where the subleading correction actually induces a significant change 
in the K\"ahler curvature. This is achieved thanks to a large numerical 
coefficient that compensates its parametrical suppression. 
We believe however that in such a situation there is limited control 
on the effect of the corrections at higher orders of the low-energy expansion.}
In models with several such fields, that is, with logarithmic potentials with 
coefficients $n_i$, one finds that the sectional curvature depends on the orientation of the Goldstino 
direction $f^i$. However one can proceed exactly as in \cite{GomezReino:2006dk} and minimise 
the sectional curvature with respect to the variables $f^i$, taking into account the constraint $f^if_i=1$. 
By doing so it is easy to find that $R(f^i) \ge 2/(\sum_i n_i)$. The condition (\ref{condalt}) 
implies therefore that $\sum_i n_i \gsim 3(1 + \gamma)$. 
As in the one field case we conclude then that one cannot get an
inflationary scale much bigger than the gravitino mass in any model 
with a small number of moduli with coefficients $n_i$ of order $1$.

Given the above two substantially different situations, one could then consider a model combining a 
field with a logarithmic K\"ahler potential and a field with a canonical K\"ahler potential (which would 
act as an uplifting sector):
\be
K = - n \log \big(T + \bar T \big) + \bar X X \,.
\ee
In such a case, the scalar manifold spanned by the fields $X$ and $T$ factorises into two 
one-dimensional manifolds. As before we find that the curvature in the one-dimensional 
manifold spanned by $X$ vanishes whereas the curvature in the one-dimensional 
manifold spanned by $T$ is given by $2/n$. This means that the minimal value that the sectional 
curvature is allowed to take is zero, since the Goldstino direction can be aligned along the direction 
of zero curvature: $R(f^i) \ge 0$. It is then always possible to satisfy the condition 
(\ref{condalt}), independently of the value of $n$. However, it is clear that in order to achieve a 
large $\gamma$, that is, a scale of inflation bigger than the gravitino scale, the inflationary dynamics 
must be strongly affected by the uplifting sector. The situation remains qualitatively the same by 
adding several such building blocks.

Another case that can be easily analysed is that of models with the following K\"ahler potential:
\be
K = - n \log \big(T + \bar T - \bar X X \big) \,.
\ee
In this case the scalar geometry is a maximally symmetric coset space with constant curvature, and 
one finds $R(f^i) = 2/n$. The situation is then identical to the one obtained with only one field $T$ 
with a logarithmic K\"ahler potential and the addition of the $X$ field does not help in satisfying the 
condition. In particular, it is impossible to realise slow-roll inflation if $n=3$, unless extra ingredients
are added.\footnote{For example the model considered in ref.~\cite{Kachru:2003sx} involves an
additional uplifting sector. In that case, besides the fields $T$ and $X$ describing the volume 
and the brane position, one would also have to take into account some extra field $Y$ describing the 
anti-brane position.} 
Again, adding more fields of this kind in a similar way does 
not change qualitatively the situation. More involved coset manifolds can be studied as in 
ref.~\cite{GomezReino:2006wv}.

\section{No-scale models}\setcounter{equation}{0}

A general feature of models emerging from string compactifications on a Calabi-Yau is that their moduli
sector exhibits, in the large volume limit, the no-scale property:
\be
K^i K_i = 3 \,.
\ee
As shown in \cite{Covi:2008ea}, this property constrains the K\"ahler geometry and, as a consequence of this,
the Riemann tensor satisfies certain properties when projected along the particular direction $k^i=K^i/\sqrt{3}$.
In particular, one finds that along such a direction the sectional curvature takes precisely the critical value:
\bea
R(k^i) = \frac 23 \,.
\eea
This means that it is always possible to obtain $\hat\sigma = 0$ by choosing the Goldstino
direction $f^i$ to be aligned along this special direction $k^i$. The question is then whether it
is possible or not, by departing from the configuration $f^{i} = k^{i}$,
to get a lower value of the sectional curvature, or equivalently, a larger value of
$\hat\sigma (f^i)$.

In orbifold models, as well as in smooth compactifications on Calabi-Yau manifolds which are actually
K3 fibrations with a large $P_1$ base,
the moduli space is a coset manifold of the type $G/H$. These spaces are symmetric and
the form of the Riemann tensor is further constrained by the presence of isometries. Moreover, they are
also homogeneous with a covariantly constant curvature. In these models, the quantity
$\hat\sigma$ can be easily studied as a function of the direction $f^i$, and it is possible to prove
that the value $\hat\sigma = 0$ along the direction $k^i$ corresponds to an absolute
maximum \cite{Covi:2008ea}. The situation is then identical to that of a single modulus with a logarithmic
potential of the form (\ref{ene}) with coefficient $n=3$:
Inflation can be realised with the help of  subleading corrections to the K\"ahler potential, but 
only with a very low scale relative to the gravitino mass ($\gamma \ll 1$). We will come back to 
this issue in the next section.

In general Calabi-Yau models, the situation is more interesting. Indeed, in those cases the scalar manifold
is in general neither symmetric nor homogeneous. The function $\hat\sigma$ can then have either a
maximum or a saddle point at the special direction $k^i$, and the space of possible
models subdivides into two classes: Models for which it is possible to find a positive value of 
$\hat\sigma$ in a
direction different than $k^i$ and models for which it is impossible to get such a positive value.

To illustrate the situation arising in more general Calabi-Yau models, let us consider the
K\"ahler moduli sector of heterotic compactifications. The K\"ahler potential is here determined by
the intersection numbers $d_{ijk}$ of the Calabi-Yau manifold and has the following form
\be\label{khet}
K = - \log \mathcal V \,, \quad \mathcal V =  \frac{1}{6} d_{ijk} (T^i + \bar T^{\bar \imath}) (T^j + \bar T^{\bar \jmath}) (T^k + \bar T^{\bar k}),
\ee
where $\mathcal V$ is the classical volume of the Calabi-Yau.
This defines a special K\"ahler geometry, and the Riemann tensor has the special structure
$R_{i \bar \jmath p \bar q} = g_{i \bar \jmath} g_{p \bar q} + g_{i \bar q} g_{p \bar \jmath}
- e^{2 K} d_{i p r} g^{r \bar s} d_{\bar s \bar \jmath \bar q}$. The sectional curvature along the Goldstino
direction is then given by:
\be
R(f^i) = 2 - e^{2K} d_{i p r} g^{r \bar s} d_{\bar s \bar \jmath \bar q} \, f^i f^{\bar \jmath} f^p f^{\bar q} \,.
\ee
This yields:
\be
\hat\sigma(f^i) = - \frac 43 + e^{2K} d_{i p r} g^{r \bar s} d_{\bar s \bar \jmath \bar q} \, f^i f^{\bar \jmath} f^p
f^{\bar q} \,.
\ee
From the explicit form of the K\"ahler metric derived from (\ref{khet}) it follows that 
$d_{ipr}k^ik^p=2k_r/\sqrt{3}$. One can then easily verify that along the special direction $k^i$ one indeed
has $R(k^i) = 2/3$ and $\hat\sigma(k^i) = 0$. It was however shown in ref.~\cite{Covi:2008ea} that $\hat\sigma$
can be made positive or negative along other directions, depending on the intersection numbers $d_{ijk}$.
For instance, in models with only two moduli, the situation simplifies due to the fact that there is only one
direction orthogonal to the direction given by $k^i$. 
This direction is given by the unit vector $n^i$ defined as:
\be
(n^1,n^2)=\frac{(k_2,-k_1)}{\sqrt{\det g}}\,,\;\;\;\;n^i\,k_i=0\,.
\ee
One can show \cite{Covi:2008ea}  that the convexity of the function $\hat \sigma (f^{i})$ at $f^{i} = k^{i}$ is determined by the
sign of the discriminant of the cubic polynomial $\mathcal V$ defining the
volume of the Calabi-Yau, given by:
\be
\Delta = - 27 \Big( d_{111}^2 d_{222}^2 - 3\, d_{112}^2 d_{122}^2 + 4\, d_{111} d_{122}^3 + 4\, d_{112}^3 d_{222}
- 6\, d_{111} d_{112} d_{122} d_{222} \Big)\,.
\ee
If $\Delta > 0$, then $\hat\sigma(k^i) = 0$ corresponds to the absolute maximum, and it is not possible
to meet the condition for slow-roll inflation. If, on the contrary, 
$\Delta < 0$, the point $\hat\sigma=0$ corresponds to a saddle point 
and therefore there is a region in the parameter space spanned by the 
$f^{i}$'s for which $\hat \sigma (f^{i})$ can be made positive.
Moreover, the value of $\hat \sigma$ for $V>0$ is extremised to a 
non-vanishing value along some particular direction $f^i$ in between 
$k^i$ and $n^i$.
Unfortunately, this value is difficult to determine in general,
essentially because $\hat \sigma$ is defined in terms of the normalised 
unit vector $f^i$.
Nevertheless, we can still verify whether it is possible or not to obtain 
$\hat \sigma(f^i)$ larger than the critical value $2/3$ required to be 
able to realise inflation with an arbitrary high scale. 
A simple way to verify that this is indeed the case is by looking at the
particular direction $f^{i} = z^{i}$ given by~\footnote{This direction was found in~\cite{Covi:2008ea} in the analysis of string compactifications with 
two moduli. There, it was shown that $z^{i}$ maximises the quantity 
$\sigma = (G^k G_k)^2 \hat\sigma (f^i)$. One should keep in mind however 
that in general the function $\hat \sigma (f^{i})$ is maximised in a 
direction $f^{i} \neq z^{i}$.}:
\be
z^i = \sqrt{\frac {1+a}{9+a}} k^i + \sqrt{\frac {8}{9+a}} n^i \,,\;\;\;\;
a = - \frac {\Delta}{24} \frac {e^{4K}}{(\det g)^3} \,.
\ee
Notice that $a>0$ as the factor $e^{4K}/(\det g)^3$ is always positive. Along this particular direction one then 
obtains\footnote{This expression can be derived as follows from the results 
of section 4 of ref.~\cite{Covi:2008ea}. One starts from the decomposition 
$\sigma = \omega - 2 s^i s_i$, with $\omega = a\, (3\, \det g \, |C|^2)^2$ 
and $s^i=0$, taking a general Goldstino direction $G_i = N_i + \alpha K_i$,
where $N_i $ is orthogonal to $K_i$. From the definition of $C$ one easily 
finds that $3\, \det g \, |C|^2 = N^i N_i$. 
Moreover, the equation $s^i=0$ fixes $\alpha$ in terms of $N^i$ and the 
arbitrary phase of $C$. One finds in particular that 
$|\alpha|^2 \ge [(1 + a)/24] N^i N_i$, the precise value depending on 
the phase of $C$. 
It then follows that $G^i G_i \ge [(a+9)/8] N^i N_i$. Finally, one 
computes $\hat\sigma = \omega/(G^iG_i)^2$, with $G^i G_i$ taken to 
assume its minimal value.}:
\be
\hat\sigma(z^i) = \frac {64 \, a}{(a + 9)^2} ,
\label{2fields}
\ee
which is positive. Then, assuming that $a$ can be varied over the whole 
range $[0,+\infty)$ by varying the values of the fields while 
keeping $e^K$, $\det g$ and ${\rm tr}\,g$ all positive, the largest possible
value for $\hat \sigma$ is obtained for $a=9$ and is given by 
$\hat \sigma_{\rm max} = 16/9$. Since this is larger than $2/3$,
one should then be able to achieve any arbitrarily large value of $\gamma$.

Another interesting situation based on Calabi-Yau manifolds arises in 
Type II orientifold compactifications.
In that case, the scalar geometry that one obtains for a given Calabi-Yau 
manifold is dual to the one arising for the heterotic model based on the 
same manifold \cite{Grimm:2004uq, D'Auria:2004cu}, and one finds opposite 
signs for the extremal value of $\hat\sigma$. 
In the special case involving only two fields, one can in fact prove that 
for orientifolds this extremal value is given by $\hat\sigma = 64 \,a/(a-9)^2$, 
where $a$ is defined as before but with $\Delta \to - \Delta$, 
$e^K \to e^{-K}$ and $\det g \to (\det g)^{-1}$, namely $a = (\Delta/24) \,e^{-4K} \,(\det g)^3$. 
In this case, a viable situation with a positive $\hat\sigma$ can therefore be realised only for
those Calabi-Yau manifolds for which $\Delta > 0$. One can actually show that in this case 
$a \in [0,1]$, and the largest possible value for $\hat \sigma$ is obtained for $a=1$ and is 
given by $\hat \sigma_{\rm max} = 1$, which is still larger than $2/3$.

\section{Effect of subleading corrections}\setcounter{equation}{0}

We would like now to discuss the role of subleading corrections in
the boundary cases when the leading order of the K\"ahler potential
just fulfills the equality in eq.~(\ref{cond}).
As we already mentioned in the last section, for no-scale models, 
the sectional curvature along 
the direction $k^i$ is $R(k^i)=2/3$, and therefore $\hat\sigma=0$ along 
that direction. This means in particular that a general possibility to 
realise inflation which can arise in all Calabi-Yau string models is to 
consider subleading corrections to the K\"ahler potential that break 
the no-scale property. However this possibility obviously restricts the 
scale of inflation to be small (compared to the gravitino scale), 
as the change in $\hat\sigma$ is of the order of the subleading correction. 

The subleading corrections to the K\"ahler potential can be of various types,
e.g. loop, $\alpha'$ or world-sheet instanton corrections. As a result of 
these corrections, the no-scale property will be deformed by some small 
quantity $\delta$, which is parametrically of order $\Delta K/K$:
\be
K^i K_i \simeq 3 + {\cal O}(\delta) \,.
\ee
In this situation, the extremum of the function $\hat\sigma$ along the direction $k^i$ gets in general
slightly shifted, and the new value at this extremum becomes of order  
\be
\hat\sigma(k^i) \simeq {\cal O}(\delta) \,.
\ee
Comparing this result with the condition (\ref{cond}), we see that in this case it would indeed be 
possible to realise inflation along the direction $f^i \simeq k^i$, provided one can get the right sign 
for the subleading correction $\delta$. However the parameter $\gamma$ which sets the scale of 
inflation is bounded by the parameter $|\delta|$ controlling the relative effect of the subleading corrections in the K\"ahler potential:
\be
\gamma \lsim {\cal O}(|\delta|) \,,
\ee
and therefore one has necessarily $H < m_{3/2}$.

One can consider for instance the effect of $\alpha'$-corrections to the large volume limit of  Calabi-Yau 
compactifications of the heterotic string~\cite{Candelas:1990rm} and of type IIB orientifolds~\cite{Becker:2002nn}. 
These corrections have the effect of shifting the argument of the logarithm in the 
K\"ahler potential by some constant parameter $\xi$~\footnote{Strictly
speaking, in the case of IIB orientifolds the correction is
dilaton-dependent. This does not qualitatively modify the effect however.}:
$$
K=-n\log(\mathcal{V}+\xi)\,,
$$
where $n=1,2$ for heterotic and orientifold models respectively. 
One can then parametrise the relative effect of these corrections with $\delta \sim \xi/\mathcal{V}$, where 
$\mathcal{V}$ is the volume of the Calabi-Yau manifold (resp. orientifold).
It is easy to check that $\hat\sigma$ still has an extremum along the 
direction $k^i$, but its value at that point becomes now 
$\hat\sigma \sim \delta \sim \xi/\mathcal{V}$. As a result, the maximal scale of inflation that can be realised within this setup 
corresponds to $\gamma \sim \xi/\mathcal{V}$, that is $H^2 \sim m_{3/2}^2 \xi/\mathcal{V}$. This is for 
example the case in the model of ref.~\cite{Conlon:2005jm}.

\section{Conclusions}\setcounter{equation}{0}

In this paper we have studied the possibility of realising successful 
slow-roll inflationary scenarios in a general low-energy effective 
supergravity theory involving only chiral multiplets. We have shown that 
the condition imposed on the theory for having slow-roll inflation is 
very similar to the one necessary for obtaining a metastable de Sitter 
vacuum. In particular, the requirement is that the sectional curvature 
$R(f^i)$ along the Goldstino direction $f^i$ should be smaller than the 
critical value $2/[3(1+\gamma)]$, where the parameter 
$\gamma= V/(3 \,m_{3/2}^2)$ depends on the size of the potential 
relative to the gravitino mass scale. As was shown in \cite{Covi:2008ea}, 
the presence of dS vacua with small cosmological constant 
$\Lambda \ll m_{3/2}^2$, that is, with $\gamma\ll 1$, implies that the 
sectional curvature is bounded, i.e. $R(f^i) \lsim 2/3$. 
For inflation, on the other hand, this condition changes depending on the 
Hubble scale. In models with $H \gg m_{3/2}$, i.e. $\gamma \gg 1$, 
such constraint becomes $R(f^i) \lsim 0$. 
For models with $H \ll m_{3/2}$ one has instead $\gamma\ll 1$ and the 
condition takes the form $R(f^i) \lsim 2/3$ and is therefore similar 
to the one relevant for metastable dS vacua. 
This means in particular that models with a scale of inflation higher 
than the gravitino mass are more difficult to realise than models with 
a scale of inflation smaller than (or comparable to) the gravitino mass.

More concretely, we have shown that the condition for successful inflation 
can be generically satisfied in any no-scale model by taking into account 
the effect of subleading corrections, although in those cases the scale 
of inflation has to be suppressed with respect to $m_{3/2}$. 
On the other hand, models with a scale of inflation that is comparable or 
even larger than the gravitino mass can instead be realised only in certain 
Calabi-Yau compactifications, those ones allowing for a value of 
$\hat\sigma\sim 2/3$. 
We have also shown through some simple examples that the conditions 
necessary for slow-roll inflation can also be achieved by adding to the 
moduli sector of the theory an uplifting sector. In those cases the size 
of the parameter $\gamma$, which gives the ratio between the scale of 
inflation and the gravitino mass during inflation, depends on the 
influence that the uplifting sector has on the inflationary dynamics. 
For example in models with a K\"ahler potential of the type (\ref{ene}) 
with $n=3$ it is clear that in order to have $\gamma\gg1$ the uplifting 
sector should dominate the inflationary dynamics. If this is not the case, 
the uplifting sector only mildly changes the condition (\ref{cond}) and 
one has a model with $H\lsim m_{3/2}$. 
This is actually the typical situation in inflationary scenarios based 
on the KKLT setup, as was pointed out in \cite{Kallosh:2004yh}. 

Recall however that the gravitino mass during inflation is not necessarily 
the same as the gravitino mass at the vacuum.  
In order to construct models with $H\gg m_{3/2}$, one possibility is then 
to perform an additional tuning to make the gravitino mass during inflation 
much bigger than the gravitino mass at the vacuum \cite{Kallosh:2004yh,jj}. 
We have shown in this paper that another possibility to realise 
$H\gg m_{3/2}$ without performing an additional tuning is to consider 
Calabi-Yau compactifications allowing for a sizable value of 
$\hat\sigma$, or equivalently, for a small value of the sectional curvature. 

It is interesting to note that from (\ref{condalt}), and by taking into 
account the definition of $\gamma$ in (\ref{gam}), one can compute the 
following bound on the value of the inflationary Hubble parameter:
\be\label{uno}
H^2 \lsim R_{\rm min}^{-1} \Big( \frac 23 - R_{\rm min} \Big) \, m_{3/2}^2 \,,
\ee
where $R_{\rm min}$ denotes the minimal value that the sectional 
curvature of the moduli space is allowed to take. 
In the vacuum of the theory the same kind of bound can be computed for the 
mass $m$ of the lightest scalar. Actually following the same reasoning as 
the one used to derive (\ref{eta-constr}) and imposing that at the vacuum 
$V=\nabla_iV=0$, one easily deduces 
that:
\be\label{dos}
m^2 \lsim\,f^i f^{\bar \jmath} \, \nabla_i\nabla_{\bar \jmath} V = 
3 \Big(\frac 23 - R_{\rm min} \Big) m_{3/2}^2 \,.
\ee
As we already mentioned, the two gravitino scales in (\ref{uno}) and 
(\ref{dos}) may differ, but in the absence of additional tuning 
of the parameters in the theory, both scales are naturally expected 
to be of the same order of magnitude. 

One can compute now the ratio of the bounds (\ref{uno}) and (\ref{dos}). 
This yields the following simple relation:
\be
\frac {H^{\max}}{m^{\rm max}} \sim R_{\rm min}^{-1/2} \,.
\ee
This is perhaps the most objective measure of the tension against making 
the scale of inflation much larger than the scale of supersymmetry breaking, 
and shows that the only way to relax such tension in a robust way 
(that is, without extra fine-tuning) is to choose for inflation a 
direction in field space where the K\"ahler curvature is very small. 

\section*{Acknowledgements}

This work was partly supported by the German Science Foundation (DFG)
under the Collaborative Research Centre (SFB) 676, by the European Union 
6th Framework Program MRTN-CT-503369 ``Quest for unification" and by the 
Swiss National Science Foundation.   
LC would like to thank  M. Badziak and M.~Olechowski for interesting
discussions and the Institute of Theoretical Physics of the 
Warsaw University for their hospitality during part of this work; 
the visit was made possible thanks to a Maria Curie Transfer of Knowledge 
Fellowship of the European Community's Sixth Framework Programme under 
contract number MTKD-CT-2005-029466 (2006-2010).
JL would like to thank Daniel Waldram for useful discussions and 
Chris Hull and the Institute for Mathematical Sciences, Imperial College 
London, for financial support and the kind hospitality during part of this
work.
GP would like to thank the Institute for Theoretical Physics of the
Ecole Polytechnique F\'ed\'erale de Lausanne for its hospitality during 
the completion of this work.

\end{document}